\documentclass[prb,twocolumn,showpacs]{revtex4}

\usepackage{epsfig,amssymb,amsmath,bm}
\begin{document}

\title{Measuring the decoherence rate in a semiconductor charge qubit}

\author{S.~D.~Barrett}\email{sean.barrett@hp.com}
\affiliation{Semiconductor Physics Group, Cavendish Laboratory,
University of Cambridge, Madingley Road, Cambridge CB3 0HE, U.K.}
\altaffiliation[Present address: ]{Hewlett-Packard Laboratories,
Filton Road, Stoke Gifford, Bristol BS34 8QZ, U.K.}

\author{G.~J.~Milburn}
\affiliation{Department of Applied Mathematics and Theoretical
Physics, University of Cambridge, Wilberforce Road, Cambridge CB3
0WA, U.K.} \affiliation{Centre for Quantum Computer Technology,
University of Queensland, St Lucia, Queensland 4072, Australia}

\date{\today}
% 03.65.Yz Decoherence; open systems; quantum statistical methods
% 03.67.Lx -Quantum computation
% 73.21.La Quantum dots
% 63.20.Kr Phonon-electron and phonon-phonon interactions

\begin{abstract}
We describe a method by which the decoherence time of a solid
state qubit may be measured. The qubit is coded in the orbital
degree of freedom of a single electron bound to a pair of donor
impurities in a semiconductor host. The qubit is manipulated by
adiabatically varying an external electric field. We show that, by
measuring the total probability of a successful qubit rotation as
a function of the control field parameters, the decoherence rate
may be determined. We estimate various system parameters,
including the decoherence rates due to electromagnetic
fluctuations and acoustic phonons. We find that, for reasonable
physical parameters, the experiment is possible with existing
technology. In particular, the use of adiabatic control fields
implies that the experiment can be performed with control
electronics with a time resolution of tens of nanoseconds.

\end{abstract}
\maketitle
\section{Introduction.}
It is generally believed that scalable quantum computing devices
will eventually be fabricated in solid state systems, and many
ideas have been
proposed.\cite{Loss1998,Kane1998,Shnirman1997,Ioffe1999,Mooij1999,
Barnes2000} Of these, only superconducting systems have, as yet,
yielded single qubit devices capable of demonstrating a large
number of single qubit rotations.\cite{Nakumura1999,Vion2002} No
multi qubit devices have been reported in the solid state. A
single qubit device however is extremely useful as it enables an
experimental measurement of the qubit decoherence time to be made.
This number will ultimately determine if a particular solid state
implementation is scalable (that is, capable of reaching the error
threshold required for fault tolerant
operation\cite{Preskill1998}). Decoherence refers to the
uncontrollable coupling between the degree of freedom coding the
qubit and other degrees of freedom in the qubit's environment.
Such uncontrollable interactions lead to the qubit becoming
entangled with these inaccessible degrees of freedom, with the
result that the state of the qubit is not precisely defined by its
preparation and subsequent control by unitary gates. Under such
circumstances, the outcomes of direct measurements on the qubit
are described by a mixed state, corresponding to an average over
the inaccessible degrees of freedom.

In solid state systems the sources of decoherence are legion, and
include phonons, nuclear spins, and electromagnetic fluctuations.
Which sources of decoherence are relevant depend on what
particular degrees of freedom are used to encode the qubit. A
great deal of experimental and theoretical work remains to be done
if we are to achieve understanding of the limitations of solid
state implementations of qubits.  In this paper we will focus on
one particular qubit encoding based on the electron charge degree
of freedom. (Decoherence of the charge degree of freedom is also
of relevance to schemes in which \emph{spin} qubits are coupled by
an exchange interaction,\cite{Loss1998,Kane1998,Barnes2000} since
charge decoherence can lead to leakage errors during exchange
interaction gates.\cite{Barrett2002})

\begin{figure}
\centerline{\epsfig{file=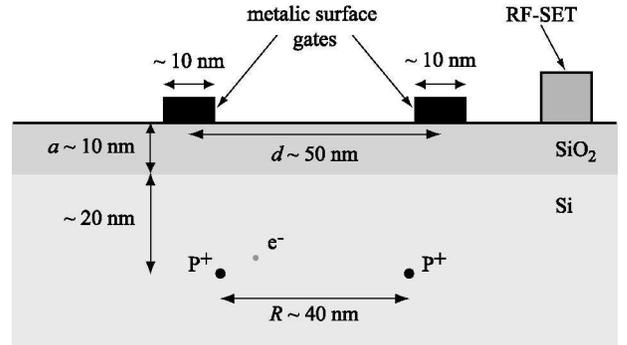}} \caption{A schematic
representation of the double donor, single electron system. The
qubit is encoded as the spatial localization of the electron
charge, relative to the donor sites. The surface gates may be used
to control the bias term, $\varepsilon$, in the qubit Hamiltonian.
The radio frequency single electron transistor (RF-SET) may be
used to readout the position of the electron.} \label{fig:device}
\end{figure}

To be specific, we will consider a system which consists of two
phosphorus donors, embedded in a silicon substrate, which share a
single excess electron.\cite{Hollenberg2003} The device is
depicted schematically in Fig. \ref{fig:device}.
The qubit is coded in terms of the relative position of the
electron. We denote localized single particle states by
$\{|L\rangle,\,|R\rangle\}$ where $|L\rangle$ corresponds to an
electron localized on the left donor site, while $|R\rangle$
denotes an electron localized on the right donor. These states are
not eigenstates of the Hamiltonian when the potential is perfectly
symmetrical. We may however represent localized states in terms of
the two lowest states of the potential; the symmetric ground state
$|E_s\rangle$ and the antisymmetric first excited state
$|E_{as}\rangle$, by $|L\,,R\rangle
=(|E_s\rangle\pm|E_{as}\rangle)/\sqrt{2}$. The system prepared in
one of these localized states will oscillate coherently between
them at the tunnelling frequency $\Delta=E_{as}-E_{s}$. If the
potential is biased sufficiently far from symmetry (by applying an
external electric field) the localized states become good
approximations to the energy eigenstates.

The single particle Hamiltonian for the double donor system may be
approximated by the two level qubit Hamiltonian
\begin{equation}
H_\textrm{qb}= -\hbar\frac{\varepsilon(t)}{2}\sigma_z
-\hbar\frac{\Delta}{2}\sigma_x \,, \label{eq:qubit}
\end{equation}
where $\sigma_x = |L\rangle \langle R| + |R\rangle \langle L|$ and
$\sigma_z = |L\rangle \langle L| - |R\rangle \langle R|$ are the
Pauli spin operators in the $\{|L\rangle,\,|R\rangle\}$ basis, and
$\varepsilon(t)$ describes the bias of the potential away from
symmetry, due to an external electric field. We have indicated
that this may be a function of time, as discussed below. For a
nonzero bias, the energy gap between the instantaneous ground
state and first excited state is $\hbar E(t) = \hbar
\sqrt{\Delta^2 + \varepsilon(t)^2}$.

This approximate Hamiltonian is valid if $\Delta$,
$|\varepsilon|$, $k_B T/\hbar \ll \omega_{01}$, where
$\omega_{01}$ is the angular frequency corresponding to
transitions between the ground and first excited states of an
electron bound to a single isolated donor. For phosphorus donors
in silicon, $\omega_{01} = 1.8 \times 10^{13}$ rad
s$^{-1}$.\cite{Pantelides1974} As we discuss in Sec.
\ref{sec:energylevels}, the tunnelling frequency,  $\Delta$,
depends on the distance between the donors. For a donor separation
of around 40 nm, the tunnelling frequency is approximately $\Delta
\sim 10^{10}$ rad s$^{-1}$.

Decoherence in this system can be due to phonons that cause
transitions between the energy eigenstates of the system. As we
show in Appendix \ref{sec:phonons}, however, the corresponding
timescale for such transitions can be made much longer than all
other timescales in the problem by choosing an appropriate donor
separation. Interactions with electromagnetic fluctuations in the
environment (e.g. due to thermal voltage noise in nearby surface
gates) however is more serious. In this paper we will model such
processes using the spin boson model. This model has been
extensively discussed in the literature (see for example Refs.
\onlinecite{Leggett1987} and  \onlinecite{Weiss1999}).

Our objective is to find a way to experimentally determine the
decoherence rate. It might be thought that this is easily done by
monitoring the decay of the coherent tunnelling oscillation, by
allowing the system to evolve for a time $t$ and then determining
the expected position of the electron relative to the double donor
system, $\langle\sigma_z\rangle(t)$. Repeating for a number of
different values of $t$, and observing the decay time of the
oscillations in $\langle\sigma_z\rangle(t)$ would yield the
decoherence rate. While this is possible in principle it is
difficult in practice, because the coherent evolution must be
turned on and off (for example, by rapidly changing the bias field
$\varepsilon(t)$), on timescales much shorter than the reciprocal
of the tunnelling frequency, $E(t)^{-1}$. Using this technique,
the tunnelling frequency itself must be much larger than the
decoherence time, which is expected to be of the order of
nanoseconds (see Sec. \ref{sec:decoherenceb}). Therefore,
measuring the decay of coherent oscillations directly would
require accurate switching of the qubit Hamiltonian on a timescale
of tens of picoseconds. Despite these difficulties, a similar
experiment has been achieved in a superconducting charge
qubit.\cite{Nakumura1999}

Other experiments have focused on \emph{continuous} measurement of
the charge degree of freedom of excess electrons in a coupled
quantum dot system, using a nearby quantum point contact
electrometer.\cite{Smith2002,Gardelis2003} A signature of charge
decoherence in the coupled dot system was observed by monitoring
the average current through the electrometer, although a large
contribution to the observed decoherence rate is thought to be due
to the back action of the electrometer on the coupled dot system.
This back action is due to the shot noise of the electrons
tunnelling through the quantum point contact.

In a recent paper,\cite{Silvestrini2000} an alternative method was
proposed to determine the decoherence rate for flux qubits
implemented in a radio frequency SQUID system.\cite{Makhlin2001}
Rather than attempting to observe the decay of coherent
oscillations of the flux, the authors proposed that the qubit
polarization be reversed by \emph{adiabatically} sweeping the
qubit Hamiltonian parameters. They argued that the decoherence
time can be determined by observing the probability of success of
the adiabatic inversion process as a function of the parameter
sweep time.

In this paper, we describe a scheme for determining the
decoherence rate in the single electron, double donor system
described above. Our scheme also makes use of adiabatic
manipulation of the Hamiltonian parameters. We show that an
experimental estimate of the decoherence rate can be obtained by
preparing the system in the ground state under strong positive
bias (a state localized on the left donor), adiabatically sweeping
the bias to zero ($\varepsilon(t)=0$) and then holding the bias at
zero for a period $t_\textrm{hold}$, before adiabatically sweeping
to the opposite bias and then determining whether or not the
system has changed its localized charge state. The final charge
state of the system can be measured using a radio frequency single
electron transistor (RF-SET).\cite{Schoelkopf1998,Aassime2001} An
RF-SET can be kept in a quiescent state during the qubit
evolution, and therefore the detector back action should not add a
significant contribution to the observed decoherence rate. A plot
of the probability of finding the electron on the right donor site
versus $t_\textrm{hold}$ will in general fall from a value close
to unity, to substantially less than unity, over a timescale
determined by the decoherence rate.

The advantages of this method over one in which coherent
oscillations are directly observed are twofold. Firstly,
substantially fewer measurements are required, since it is not
necessary to plot out several coherent oscillations. Secondly, the
timescales over which $\varepsilon(t)$ must be varied are
determined by the decoherence timescale itself, rather than the
(much shorter) timescale for coherent oscillations, $E(t)^{-1}$.
Estimates of the relevant parameters, presented below, suggest
that the experiment can be performed with electronics with a time
resolution of tens of nanoseconds, rather than tens of
picoseconds.

This paper is organized as follows. In Sec. \ref{sec:energylevels}
we estimate the tunnelling frequency, $\Delta$, for the double
donor, single electron system, as a function of the donor
separation. In Sec. \ref{sec:outline} we describe the scheme for
determining the decoherence rate in more detail. In Sec.
\ref{sec:decoherencea} introduce the spin-boson model for the
coupling of the qubit to the environment. In Sec.
\ref{sec:decoherenceb} we calculate an estimate for the strength
of the system-environment coupling for the case of decoherence due
to thermal voltage noise in nearby surface gates, and present the
results of numerical calculations of the evolution of the qubit
under such a coupling. In order for the experiment to be viable, a
number of constraints must be satisfied. We quantitatively discuss
these in Sec. \ref{sec:discussion}, and also find a set of
experimentally achievable parameters that satisfy these
constraints. We also discuss a number of other issues relating to
the implementation of this scheme.

\section{Approximate energy levels of the single electron,
double donor system} \label{sec:energylevels}

\begin{figure}
\centerline{\epsfig{file=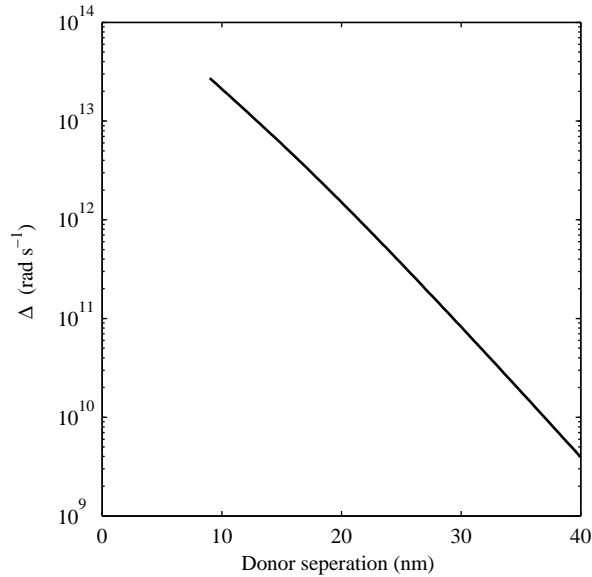}} \caption{Approximate energy
gap, $\Delta$, between the lowest symmetric and antisymmetric
eigenstates of double donor, single electron system, under zero
bias, as a function of donor separation.} \label{fig:energylevels}
\end{figure}
%
% sefig2.eps plotted with H2+\plotdelta.m
%
%

The tunnelling frequency, $\Delta$, may be estimated by
determining approximate energy eigenvalues for the lowest energy
symmetric and antisymmetric eigenstates for the double donor,
single electron system. Finding exact values for these energy
levels is complicated by the fact that the conduction band
electron dispersion relation in silicon is anisotropic, and also
by the valley-orbit interaction.\cite{Yu1996} For the purposes of
this work, however, it will be sufficient to gain an order of
magnitude estimate for $\Delta$. To this end, we ignore the
conduction band anisotropy and assume that the localized states
$|L\rangle$ and $|R\rangle$ may be represented by 1s-orbitals
centered on the left and right donor sites respectively. We take
the Bohr radius for an isolated donor state to be
\begin{equation}
a_B^* = \epsilon_\textrm{Si} \left(\frac{m_e}{m_T}\right) a_B\,,
\end{equation}
where $\epsilon_\textrm{Si} = 11.7$ is the dielectric constant for
silicon, $m_e$ is the mass of a free electron, $m_T = 0.2 m_e$ is
the transverse conduction band effective mass in silicon, and $a_B
= 5.3 \times 10^{-11}$ m is the Bohr radius for the hydrogen
atom.\cite{nsm} We take the binding energy of a single electron to
a single donor to be the experimentally observed value of -45.5
meV.\cite{Pantelides1974}

With these assumptions, the energy levels of the double donor,
single electron system can be determined by the same variational
linear combination of atomic orbitals (LCAO) technique used to
calculate the eigenvalues of an H$_2^+$ molecule.\cite{Slater1965}
In contrast to a real H$_2^+$ molecule, however, the position of
the donors is fixed within the silicon lattice, and so it is not
necessary to minimize the energy with respect to the donor
separation. We plot the tunnelling frequency, $\Delta$, as a
function of the donor separation, $R$, in Fig.
\ref{fig:energylevels}.

\section{Outline of the scheme}
\label{sec:outline}

The scheme for measuring the decoherence rate for the charge qubit
proceeds as follows. Initially, the electron is prepared in the
$|L\rangle$ state by placing a large electric field across the
double donor system, such that the bias term in the qubit
Hamiltonian takes the value $\varepsilon_0 \gg \Delta$.
$\varepsilon_0$ must be chosen such that the total energy gap for
the two level system satisfies $\hbar E = \hbar \sqrt{\Delta^2 +
\varepsilon_0^2} > k_B T$. The electron will then relax to the
ground state, which is strongly localized on the left donor site.

The system is then placed in the symmetric superposition state,
$(|L\rangle + |R\rangle)/\sqrt{2}$, by adiabatically sweeping the
bias field to the symmetry point $\varepsilon(t) = 0$. The bias
field sweep should be performed quickly, so that there is
negligible decoherence during the sweep. However the sweep must
not be made too quickly, or there will be coherent non-adiabatic
transitions into the excited state. We discuss these requirements
in more detail in Sec. \ref{sec:discussion}.

The bias field is held at zero for a time $t_\textrm{hold}$.
During this time, as a result of the interaction with the
environment, the qubit will lose coherence. This loss of coherence
will be manifest in the decay of the off diagonal elements of the
qubit density matrix (written in the ${|L\rangle,\,|R\rangle}$
basis) at a rate $\Gamma_0$.

The bias field is then swept, rapidly but adiabatically, to a
large negative value, $-\varepsilon_0$, and held at this value
while the position of the electron is read out by a nearby
electrometer. Thus $\varepsilon(t)$ has the following time
dependence:
\begin{equation}
\varepsilon(t) = \left \{
\begin{array}{lrl}
\varepsilon_0 & &t \le 0 \\
\varepsilon_0 \frac{t_\textrm{sw} - t}{t_\textrm{sw}} & 0 <
&t \le t_\textrm{sw} \\
0 & t_\textrm{sw} < &t \le t_\textrm{sw} + t_\textrm{hold} \\
-\varepsilon_0 \frac{t - t_\textrm{sw} -
t_\textrm{hold}}{t_\textrm{sw}} & t_\textrm{sw} +
t_\textrm{hold} < &t \le 2 t_\textrm{sw} + t_\textrm{hold} \\
-\varepsilon_0 & 2 t_\textrm{sw} + t_\textrm{hold} < &t \,,
\end{array} \right.
\end{equation}
where $t_\textrm{sw}$ is the time taken for each bias sweep.

Provided the electrometer can determine the position of the
electron charge in a timescale shorter than the relaxation time
for the qubit under the large bias ($-\epsilon_0$), the readout
process will correspond to a strong quantum measurement in the
$\{|L\rangle,\,|R\rangle\}$ basis. As we discuss further in Sec.
\ref{sec:discussion} this measurement can be implemented by
existing radio frequency single electron transistor (RF-SET)
technology.\cite{Schoelkopf1998,Aassime2001,Buehler2002}

By repeating the above preparation, bias sweep, and measurement
steps a number of times, one can determine the probability, $P_R$,
of finding the electron on the right donor site at the end of the
sweep. If the decay of the off-diagonal elements of the density
matrix (in the ${|L\rangle,\,|R\rangle}$ basis) during the time
$t_\textrm{hold}$ for which the bias is held at $\varepsilon=0$ is
negligible, then the electron will coherently tunnel through to
the right donor site as the bias is swept through to
$\varepsilon=-\varepsilon_0$. The final state will be,
approximately, the pure state, $|R\rangle$. Thus the observed
probability of finding the electron on the right donor will be
close to unity. Conversely, if there has been substantial decay of
the off-diagonal elements during the `hold' part of the evolution,
the final state will be mixed, and the observed $P_R$ will be
substantially less than unity. The off-diagonal density matrix
elements are expected to decay over a timescale $\Gamma_0^{-1}$.
Thus repeating the whole procedure for different values of
$t_\textrm{hold}$, and plotting $P_R$ as a function of
$t_\textrm{hold}$ should allow one to determine $\Gamma_0^{-1}$.

\section{Estimating the decoherence rate by adiabatic tunnelling}
\label{sec:decoherence}
\subsection{The model of decoherence}
\label{sec:decoherencea}
In order to study the effects of the
environment on the qubit, we model the environment as a bath of
harmonic oscillator modes, linearly coupled to the $z-$component
of the qubit, via the spin boson Hamiltonian,
\begin{equation}
H = H_\textrm{qb} + \sigma_z \sum_i \hbar \lambda_i (a^{\dagger}_i
+ a_i) + \sum_i \hbar \omega_i a_i^\dagger a_i \,.\label{eq:SB}
\end{equation}
$H_\textrm{qb}$ is the qubit Hamiltonian of Eq.(\ref{eq:qubit}).
The second term describes the coupling between the position degree
of freedom of the electron ($\sigma_z$), and the displacement
operators for the bath modes ($a^{\dagger}_i + a_i$), where the
$\lambda_i$'s are coupling coefficients. The last term represents
the free Hamiltonian of the oscillator bath, where the
$\omega_i$'s are the angular frequencies of the bath modes.

The spin boson Hamiltonian has been studied
extensively.\cite{Weiss1999,Leggett1987} The behavior of the
system depends crucially on the \emph{spectral density} of the
bath, defined as
\begin{equation}
J(\omega) = \sum_i \lambda_i^2 \delta(\omega-\omega_i)\,.
\label{eq:spectral}
\end{equation}
In general, the dynamics of the spin-boson model, for an arbitrary
spectral density, is rather complicated. For the purposes of this
work, however, a number of simplifying assumptions can be made.
Firstly, we assume that the sweep of the bias field,
$\varepsilon(t)$, is made sufficiently slowly that an adiabatic
approximation can be employed. In particular, we
require\cite{Zener1932}
\begin{equation} \frac{\pi}{2}\frac{\Delta^2
t_\textrm{sw}}{\varepsilon_0} \gg 1 \,. \label{eq:adiabaticity}
\end{equation}
(We discuss this adiabaticity requirement further in Sec.
\ref{sec:discussion}.) Secondly, we assume a weak system-bath
coupling, such that $J(k_\textrm{B}T/\hbar) \ll E(t)$, throughout
the sweep. Finally, we take the initial state of the qubit to be
the thermal state
\begin{equation}
\rho_0 = \frac{\exp(-H_{\textrm{qb},0}/k_B T)}
{\textrm{tr}[\exp(-H_{\textrm{qb},0}/k_B T)]} \,,
\label{eq:initialState}
\end{equation}
where $H_{\textrm{qb},0}$ is the initial qubit Hamiltonian, i.e.
Eqn. (\ref{eq:qubit}) with $\varepsilon(t)=\varepsilon_0$. Note
that $\rho_0$ is diagonal in the energy eigenbasis of the initial
qubit Hamiltonian. Under these assumptions, the density matrix of
the qubit is always diagonal in the instantaneous energy
eigenbasis of the qubit Hamiltonian.\cite{Stockburger1995} In this
case, the Bloch vector $\vec{r}(t) = \left(\langle\sigma_x\rangle,
\langle\sigma_y\rangle,\langle\sigma_z\rangle\right)$ always lies
parallel to the vector $\vec{B} =
\left(\Delta,0,\varepsilon(t)\right)$, and the dynamics can be
understood by considering the evolution of $r(t) =
\left|\vec{r}(t)\right|$, the length of the Bloch vector. The
evolution of $r(t)$ under the above assumptions, is given
by\cite{Stockburger1995}
\begin{equation}
\dot{r}(t) = -\Gamma(t) \left(r(t) -
r_\textrm{eq}(t)\right)\,,\label{eq:rdot}
\end{equation}
where the instantaneous relaxation rate $\Gamma(t)$ depends on the
spin boson model parameters,\cite{Weiss1999,Leggett1987}
\begin{equation}
\Gamma(t) = \frac{\pi}{2} \sin^2\theta J\left(E(t)\right)
\coth\left(\frac{\hbar E(t)}{2k_\textrm{B}T}\right)
\,,\label{eq:Gamma}
\end{equation}
where $\theta = \tan^{-1}(\Delta/\varepsilon)$. $r_\textrm{eq}(t)$
is the thermal equilibrium value of the Bloch vector, evaluated
for the instantaneous energy gap of the system, $r_\textrm{eq}(t)
= \tanh\left(\hbar E(t)/2k_\textrm{B}T\right)$.

At low frequencies, the spectral density of the bath typically has
a power law behavior,\cite{Weiss1999,Leggett1987} $J(\omega)
\propto \omega^s$, where the exponent $s$ depends on the nature of
the environment. Two potentially serious sources of decoherence in
this system are a deformation potential coupling between the qubit
and acoustic phonons, and an electrostatic coupling to
Nyquist-Johnson voltage fluctuations, which may originate in the
surface electrodes used to control the qubit Hamiltonian
parameters. The former is described by a superohmic spectral
density ($s>1$). However, as we show in Appendix
\ref{sec:phonons}, with a judicious choice of donor configuration,
the decoherence rate due to phonons can be made negligibly small,
and therefore we neglect it in what follows.

\subsection{Results for Ohmic damping}
\label{sec:decoherenceb}

In this section, we concentrate on the case of decoherence due to
Nyquist-Johnson voltage noise, which is characterized by a bath
with an Ohmic spectral density ($s=1$). At low frequencies, the
spectral density may be written\cite{Weiss1999,Leggett1987}
\begin{equation}
J(\omega) = 2 \alpha \omega \,, \label{eq:JOhmic}
\end{equation}
where $\alpha$ is a dimensionless parameter which characterizes
the strength of the system-bath coupling.

In order to estimate $\alpha$, we follow a similar procedure to
that employed in Ref. \onlinecite{Makhlin2001}. We first define
the bath operator
\begin{equation}
X =  \sum_i \lambda_i \left( a^\dag_i + a^{\phantom \dag}_i
\right) \,, \label{eq:X}
\end{equation}
which couples to the $\sigma_z$ operator of the qubit, via the
second term in Eq.(\ref{eq:SB}). To proceed, we calculate the
spectrum of fluctuations in $X$ in terms of the spectral density
$J(\omega)$, and relate this to the spectrum of Nyquist-Johnson
fluctuations in the surface gates. For a bath of harmonic
oscillator modes in thermal equilibrium at temperature $T$, the
Fourier transform of the symmetrized correlation function of this
operator takes the form
\begin{eqnarray}
S_X(\omega) & = & \int_{-\infty}^{\infty}
\frac{1}{2}\left\langle\left[ X(t+\tau),
X(t)\right]_+\right\rangle e^{-i \omega \tau} d \tau
\nonumber \\
&=& \pi J(\omega) \coth\left(\frac{\hbar\omega}{2 k_B T}\right)\,,
\label{eq:corrX}
\end{eqnarray}
where $[A,B]_+ = AB+BA$ denotes an anti-commutator, $X(t) = e^{i H
t /\hbar} X e^{-i H t /\hbar}$ is the bath operator in the
Heisenberg picture, and $\langle O \rangle = \textrm{tr} [O
\rho_\textrm{env}]$ denotes the expectation of $O$ for an
environment in a thermal equilibrium state, $\rho_\textrm{env}$.

For noise due to voltage fluctuations, $X$ may be related to a
perturbation $\delta V_{LR}$ in the potential difference between
the two donor sites by
\begin{equation}
X = \frac{e\delta V_{LR}}{2 \hbar}\,. \label{eq:XdeltaV}
\end{equation}
$\delta V_{LR}$ is related to the voltage fluctuations in the
surface gates by
\begin{equation}
\delta V_{LR} = \beta \delta V_{\textrm{gate}}\,.
\label{eq:betadef}
\end{equation}
where the dimensionless parameter $\beta$ quantifies the
electrostatic coupling between the surface gates and the donor
sites, and is determined by the device geometry. For the geometry
shown in Fig. \ref{fig:device}, $\beta$ may be approximated by
elementary electrostatics as
\begin{equation}
\beta \approx \frac{2\ln\left(\frac{r_2}{r_1}\right)}{\left\lbrace
1+\frac{\epsilon_2}{\epsilon_1}\right\rbrace
\left\lbrace\ln\left(\frac{d-r_0}{r_0}\right) +\frac{1}{2}
\left(\frac{\epsilon_1-\epsilon_2}{\epsilon_1+\epsilon_2}\right)
\ln \left(\frac{(d-r_0)^2 + 4a^2}{r_0^2+4a^2}\right)\right\rbrace}
\,, \label{eq:beta}
\end{equation}
where $d$ is the distance between the two surface electrodes, $a$
is the thickness of the oxide layer, $r_1$ is the distance between
the left donor and the left electrode, $r_2$ is the distance
between the left donor and the right electrode, $r_0$ is the
effective radius of the electrode, and $\epsilon_1$ and
$\epsilon_2$ are the dielectric constants of the oxide and silicon
layers, respectively. In deriving this expression we have assumed
that the gates may be represented by long, cylindrical conductors,
and that $r_0 \ll a, d$. Using the values for $r_1$, $r_2$, $d$,
and $a$ given in Fig. \ref{fig:device}, and taking $r_0 = 5\,$nm,
$\epsilon_1 = 4$, and $\epsilon_2 = 12$, we find $\beta = 0.17$.

Substituting Eq.(\ref{eq:betadef}) into Eq.(\ref{eq:XdeltaV}), and
calculating the corresponding power spectrum yields
\begin{equation}
S_X(\omega) = \frac{e^2 \beta^2}{4 \hbar^2} S_V(\omega)\,.
\label{eq:corrX2}
\end{equation}
For Nyquist-Johnson noise, the voltage fluctuations are
characterized by\cite{Callen1951}
\begin{eqnarray}
S_V(\omega) &=& \int_{-\infty}^{\infty} \left\langle \delta
V_\textrm{gate}(t+\tau) \delta V_\textrm{gate}(t)\right\rangle
e^{-i \omega \tau} d \tau
\nonumber \\
&=& R_\textrm{gate} \hbar \omega \coth \left(\frac{\hbar \omega}{2
k_B T} \right)\,, \label{eq:NJ}
\end{eqnarray}
where $R_\textrm{gate}$ is the impedance of the circuit which
generates the gate voltages, and $T$ is the corresponding noise
temperature. Substituting this expression into Eq.
(\ref{eq:corrX2}) and comparing with Eqs. (\ref{eq:JOhmic}) and
(\ref{eq:corrX}), we find that the system-bath coupling parameter
is
\begin{equation}
\alpha = \frac{\beta^2 R_\textrm{gate} }{4R_Q}\,, \label{eq:alpha}
\end{equation}
where $R_Q = h/e^2 = 25.8 \, \textrm{k}\Omega$ is the quantum
resistance. Taking $R_\textrm{gate} = 50 \, \Omega$ and $\beta =
0.17$, we have $\alpha = 1.4 \times 10^{-5}$.

\begin{figure}
\centerline{\epsfig{file=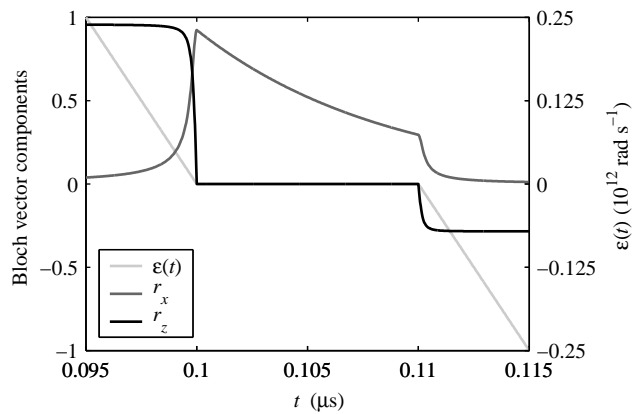}} \caption{Sample
evolution of the Bloch vector components $r_x$ and $r_z$, for part
of the bias sweep. Also shown is the time profile of the bias
sweep itself, $\varepsilon(t)$ (right axis).}
\label{fig:sampleevolutions}
\end{figure}
%
% sampleevolutions.ps plotted with H2+\sweep\cus files\sampleevolfig.m
%
%

We numerically solved Eqs. (\ref{eq:rdot}) and (\ref{eq:Gamma})
for the Ohmic spectral density of Eq. (\ref{eq:JOhmic}), assuming
a bias sweep $\varepsilon(t)$ of the form described in section
\ref{sec:outline}. Figure \ref{fig:sampleevolutions} shows the
evolution of the $x$ and $z$ components of the Bloch vector,
$\vec{r}(t)$, for a bias sweep with parameters $\varepsilon_0 = 5
\times 10^{12} \, \textrm{s}^{-1}$, $t_\textrm{sw}= 10^{-7} \,
\textrm{s}$, and $t_\textrm{hold}=10^{-8} \, \textrm{s}$. We
assume also that $\Delta = 10^{10} \, \textrm{s}^{-1}$ and $T = 10
\, \textrm{K}$.

\begin{figure}
\centerline{\epsfig{file=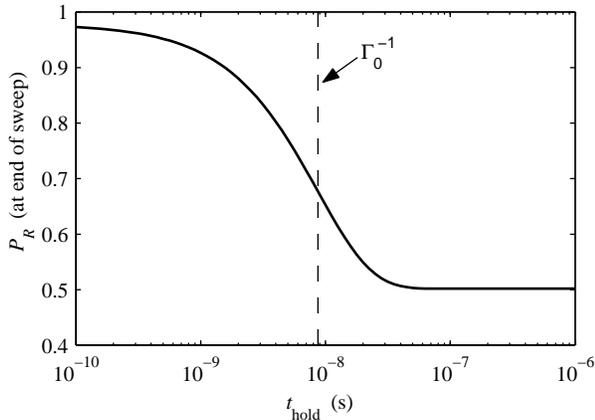}} \caption{Probability of finding
electron on the right donor, $P_R$, at the end of the bias sweep,
as a function of $t_\textrm{hold}$. The broken line represents the
reciprocal of the decoherence rate for zero bias,
$\Gamma_0^{-1}$.} \label{fig:pt}
\end{figure}

Figure \ref{fig:pt} shows the resultant probability, $P_R =
(1-r_z)/2$, that the electron is found on the right donor at the
end of the sweep, as a function of $t_\textrm{hold}$. The other
parameters used in this calculation are the same as those used in
Fig. \ref{fig:sampleevolutions}. For values of $t_\textrm{hold}
\ll \Gamma_0^{-1}$, $P_R$ is close to unity, indicating that the
electron has coherently tunnelled from the left donor site to the
right donor site. Note that $P_R$ saturates to a value slightly
less than unity, as a result of a small amount of decoherence
during the `sweep' parts of the evolution. For values of
$t_\textrm{hold} \gtrsim \Gamma_0^{-1}$, the resultant probability
is substantially less than unity, indicating a loss of coherence
during the `hold' part of the evolution, due to interaction with
the environment. The transition between these regimes occurs at a
value of $t_\textrm{hold} \sim \Gamma_0^{-1}$. Thus measuring
$P_R$ at the end of the sweep provides a method for estimating the
decoherence time $\Gamma_0$, and hence for estimating the strength
of the system-environment coupling.

%
% pl.eps plotted with H2+\sweep\cus files\manyholds.m
% \Gamma_0 label and arrow added by hand
%

\section{Discussion}
\label{sec:discussion}

In order that the transition from coherent tunnelling ($P_R
\approx 1$ at the end of the sweep) to incoherent behavior ($P_R$
substantially less than $1$) can be observed, and hence $\Gamma_0$
be determined, the parameters $\Delta$, $\varepsilon_0$ and
$t_\textrm{sw}$ must satisfy a number of constraints. Firstly, we
require that at time $t=0$, the electron must be strongly
localized at the left hand donor site. This can be achieved by
placing a large bias $\varepsilon_0$ across the double donor
system, and waiting for the donor to relax to its ground state.
This implies that we require
\begin{equation}
\hbar \varepsilon_0 \gg k_B T \,. \label{eq:localise}
\end{equation}

Secondly, we require that the minimum energy gap between the
ground and excited states satisfies
\begin{equation}
\hbar \Delta \ll k_B T \,, \label{eq:deltasmall}
\end{equation}
otherwise the system will simply remain in its ground state,
throughout the bias sweep, and it will not be possible to observe
the effects of decoherence.

Thirdly, coherent, non-adiabatic transitions into the excited
level should be minimized. The problem of non-adiabatic
transitions in two level systems was considered by
Landau\cite{Landau1932} and Zener\cite{Zener1932}. The results of
Ref. \onlinecite{Zener1932} are directly applicable to the present
work. For negligible nonadiabatic transitions, we require
\begin{equation}
\frac{\pi}{2}\frac{\Delta^2 t_\textrm{sw}}{\varepsilon_0} \gg 1
\,. \label{eq:adiabaticity2}
\end{equation}

Finally, it is necessary to ensure that relaxation at the end of
the bias sweep (when $\varepsilon(t) < -k_B T$) is negligible. If
there is significant relaxation over the last part of the bias
sweep or during the measurement process, the system will be found
to be in its ground state regardless of $t_\textrm{hold}$, and it
will not be possible to observe the effects of decoherence. The
probability that the electron will relax into the ground state,
over the last part of the sweep, is approximately
\begin{equation}
P_\textrm{relax} \approx \int_{t_*}^{t_f} \Gamma(t) dt \,.
\label{eq:relaxProb}
\end{equation}
where $\Gamma(t)$is the relaxation rate of Eq. (\ref{eq:Gamma}),
and $t_*$ is the time for which $\varepsilon(t) = \varepsilon_* =
-k_B T$, and $t_f = 2 t_\textrm{sw}+t_\textrm{hold}$ is the time
corresponding to the end of the sweep. Performing the integral,
and requiring that $P_\textrm{relax} \ll 1$, we have
\begin{equation}
\frac{\alpha \pi \Delta^2}{\varepsilon_0}
\ln\left(\frac{\varepsilon_0}{\varepsilon_*}\right) \ll
t_\textrm{sw}^{-1}\,. \label{eq:relax}
\end{equation}
In arriving at this expression, we have made the approximation
$\coth(\hbar E(t) / 2 k_B T) \approx 1$ for $t \ge t_*$, and that
$\varepsilon_0 \gg \varepsilon_*$.

In order that there is no significant relaxation to the ground
state during the measurement process, we require
\begin{equation}
\frac{\alpha \pi \Delta^2}{\varepsilon_0}
 \ll
t_\textrm{meas}^{-1}\,, \label{eq:relaxMeas}
\end{equation}
where $t_\textrm{meas}$ is the characteristic time for the
electrometer to detect the presence or absence of the electron on
the right donor site. We take $t_\textrm{meas} = 1 \mu s$, which
is readily achievable with existing radio frequency single
electron transistor (RF-SET)
technology.\cite{Schoelkopf1998,Aassime2001,Buehler2002}

In the preceding analysis, $T$ corresponds to the noise
temperature of the electronics which generate the bias sweep.
Taking $T = 10 \, \textrm{K}$, we find that $k_B T/\hbar = 1.3
\times 10^{12} \, \textrm{s}^{-1}$. In order to satisfy the
inequalities of Eq. (\ref{eq:localise}) and Eq.
(\ref{eq:deltasmall}), we choose $\Delta = 10^{10} \,
\textrm{s}^{-1}$ and $\varepsilon_0 = 5 \times 10^{12} \,
\textrm{s}^{-1}$. The inequality of Eq. (\ref{eq:adiabaticity2})
can then be satisfied if we choose $t_\textrm{sw} = 10^{-7} \,
\textrm{s}$. With these parameters, Eq. (\ref{eq:relax}) and Eq.
(\ref{eq:relaxMeas}) imply that unwanted relaxation is negligible
provided $\alpha \le 3 \times 10^{-3}$. Comparison with our
earlier estimate, from Sec. \ref{sec:decoherenceb}, of $\alpha
\approx 10^{-5}$, suggests that the experiment is indeed feasible.

A central element of the scheme introduced in Sec.
\ref{sec:outline} is that the bias field, $\varepsilon(t)$, is
held at zero for a time $t_\textrm{hold}$. This bias field will be
related to $V_\textrm{gate}$, the voltage across the surface
electrodes in Fig. \ref{fig:device}. However, imperfections in the
fabrication of a real device, and the existence of other surface
electrodes (for instance, the plunger gate used to tune the
RF-SET), may alter the potential landscape in the vicinity of the
donors, leading to a small systematic error, $\delta \varepsilon$,
in the bias field. This will lead to a systematic error in the
observed value of the decoherence rate. According to Eq.
(\ref{eq:Gamma}), provided $\hbar \delta\varepsilon \ll k_B T$,
the observed rate will be
\begin{equation}
\Gamma'_0 = \frac{\Delta^2}{\Delta^2 + \delta\varepsilon^2}
\Gamma_0 \,, \label{eq:GammaErr}
\end{equation}
where $\Gamma_0$ is the decoherence rate evaluated for $\delta
\varepsilon = 0$. Thus the true decoherence rate can be inferred
by determining $\Gamma'_0$ for a range of different offset
voltages and fitting the results to Eq. (\ref{eq:GammaErr}). Note
that for sufficiently small offsets, $\delta \varepsilon <
\Delta$, we have
\begin{equation}
\Gamma'_0 \approx \left(1 -
\frac{\delta\varepsilon^2}{\Delta^2}\right) \Gamma_0 \,,
\label{eq:GammaErrApprox}
\end{equation}
i.e. the error in the observed decoherence rate is only quadratic
in the offset error.

In our discussion of decoherence mechanisms, we have not
explicitly considered errors due to background charge
fluctuations. These fluctuations vary from sample to sample, and
typically have a $1/f$ spectrum with a shoulder at $100 - 1000$
Hz.\cite{Zorin1995} This timescale is longer than the time taken
for each preparation, sweep, and measure cycle. Background charge
fluctuations will therefore have the same effect as adding a small
random offset bias, $\delta \varepsilon$, which may vary between
cycles, but will be essentially constant over each bias sweep. As
described above, the effect of such an offset becomes unimportant
provided that, in a given sample, the charge fluctuations are
sufficiently small that the corresponding offsets satisfy $\delta
\varepsilon < \Delta$.

\section{Conclusion}
\label{sec:conclusion}

In summary, we have proposed and analyzed, theoretically, an
experimentally feasible scheme for directly determining the
decoherence rate for a solid state charge qubit consisting of a
single electron bound to a pair of donor impurities in a
semiconductor host. The qubit is manipulated by adiabatically
varying the bias term in the Hamiltonian. For a specific
implementation using phosphorous donors embedded in a silicon
host, we have theoretically obtained quantitative estimates for
the Hamiltonian parameters, and for decoherence rates
corresponding to interactions with both acoustic phonons and
voltage fluctuations. We have analyzed various constraints which
must be satisfied in order for the experiment to be feasible. We
have found appropriate, experimentally achievable parameters which
satisfy these constraints. Our results indicate that the control
field needs to be manipulated with a time resolution of tens of
nanoseconds, which is well within reach of current technology.
Performing this experiment would be a vital step towards the
implementation of a scalable solid state quantum computer.

\begin{acknowledgments}
GJM acknowledges the support of the CMI at the Department of
Applied Mathematics and Theoretical Physics, University of
Cambridge. SDB thanks the EPSRC for financial support. We would
like to thank Alex Hamilton, Tom Stace, and Cameron Wellard  for
valuable discussions. In addition we thank Crispin Barnes, Richard
George, Charles Smith, and Tim Spiller for useful comments
regarding this manuscript.
\end{acknowledgments}

\appendix
\section{Relaxation due to phonons}
\label{sec:phonons}

In this appendix, we estimate the decoherence rate due to
interaction with acoustic phonons. The problem of electron
scattering by acoustic phonons in silicon was originally
considered by Bardeen and Shockley.\cite{Bardeen1950} More
recently, electron relaxation, due to phonons, in low dimensional
semiconductor systems has been
considered.\cite{Bockelmann1990,Benisty1995} Due to the
confinement of the electrons in these systems, and the resulting
discrete spectrum of the electronic energy levels, relaxation due
to phonons is suppressed.

The rate for phonon emission in confined systems
is\cite{Bockelmann1990,Benisty1995}
\begin{eqnarray}
\Gamma_{\textrm{ph}} &=& \frac{D^2 q_{if}^3}{8 \pi^2\rho\hbar
c_s^2}\left[ n_B(E,T_\textrm{ph})+1 \right] \nonumber\\
&& \times \int{}{}d \Omega_{q} |\langle \psi_f |
e^{i\vec{q}\cdot\vec{r}}| \psi_i \rangle |^2
\,,\label{eq:FGRphonon}
\end{eqnarray}
where $D$ is a deformation potential, $\rho$ is the density of
silicon, $c_s$ is the speed of sound, $\hbar E$ is the energy
difference between the initial and final electron states, $n_B
(\omega,T_\textrm{ph}) = (\exp(\hbar \omega / k_B
T_\textrm{ph})-1)^{-1}$ is the Bose occupation function for a bath
of phonons at temperature $T_\textrm{ph}$, and $q_{if}$ is the
wave number of the emitted phonon. $q_{if}$ is fixed by the energy
gap between the ground and excited states and the phonon
dispersion relation as $q_{if} = E / c_s$. The integral in Eq.
(\ref{eq:FGRphonon}) is over all solid angles in $q$ space, and is
evaluated for $q = q_{if}$. In general, owing to the anisotropy of
the crystal, both $D$ and $c_s$ will be tensors. However, for the
purpose of gaining an order of magnitude estimate of
$\Gamma_\textrm{ph}$, we will ignore these subtleties and treat
these quantities as being isotropic.

\begin{figure}
\centerline{\epsfig{file=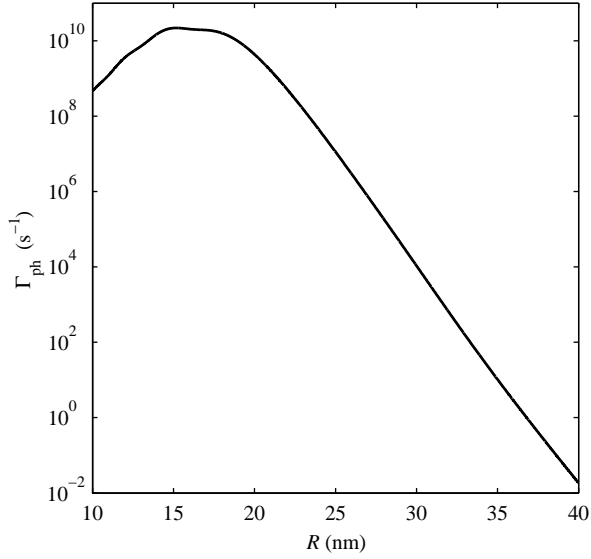}} \caption{Approximate
relaxation rate due to LA phonons, $\Gamma_\textrm{ph}$, as a
function of donor separation, for zero bias ($\varepsilon = 0$).
We assume $T_\textrm{ph} = 0.1$ K, $D = 3.3$ eV, $\rho = 2.33
\textrm{ kg m}^{-3}$, $c_s = 9.0 \times 10^3 \textrm{m s }^{-1}$
and $a_0 = 3 \textrm{ nm}$.} \label{fig:phononzerobias}
\end{figure}
%
% phonons.eps plotted with H2+\plotrelaxfig.m
%

The initial and final electron states are
\begin{equation}
| \psi_i \rangle = \cos\frac{\theta}{2} | L \rangle +
\sin\frac{\theta}{2} | R \rangle \,,
\end{equation}
\begin{equation}
| \psi_f \rangle = \sin\frac{\theta}{2} | L \rangle -
\cos\frac{\theta}{2} | R \rangle \,,
\end{equation}
where $\theta = \tan^{-1}(\Delta/\varepsilon)$, and
$ | L \rangle = (\pi a_{B}^3)^{-\frac{1}{2}} \exp(-r_a/a_{B})$
and $ | R \rangle = (\pi a_{B}^3)^{-\frac{1}{2}} \exp(-r_b/a_{B})$
are 1s-orbitals, with Bohr radius $a_B$, localized on the left and
right donor site respectively. Taking the origin to be the mid
point of the line joining the two donors, we have $\vec{r}_a =
\vec{r}+\frac{1}{2}R\vec{u}_x$ and $\vec{r}_b =
\vec{r}-\frac{1}{2}R\vec{u}_x$, where $\vec{u}_x$ is the unit
vector along the line joining the two donors, and $R$ is the donor
separation. In these coordinates, the matrix element in Eq.
(\ref{eq:FGRphonon}) may be written
\begin{eqnarray}
\langle \psi_f | e^{i\vec{q}\cdot\vec{r}}| \psi_i \rangle & =&
\frac{\sin \theta }{ 2 \pi a_{B}^3} \int{}{} dV
e^{i\vec{q}\cdot\vec{r}} \left(e^{-2r_a /a_{B}} - e^{-2r_b
/a_{B}} \right) \nonumber \\
&& + O\left( (R/a_{B})^3 e^{-R /a_{B}} \right) \,.
\end{eqnarray}
The last term in this expression may be neglected for donor
separations $R \gg 3 a_{B}$. Performing this integral, with the
aid of the convolution theorem, we find that
\begin{equation}
\langle \psi_f | e^{i\vec{q}\cdot\vec{r}}| \psi_i \rangle =
\frac{-16 i \sin \theta \sin(q_x R/2)}{\left[(q a_{B})^2 + 4
\right] ^2}\,,
\end{equation}
where $q_x$ is the component of the phonon wavevector along the
line joining the two donors. Substituting this expression into Eq.
(\ref{eq:FGRphonon}), and performing the integral over all solid
angles, gives
\begin{equation}
\Gamma_{\textrm{ph}} = \frac{64 D^2 q_{if}^3 \sin^2 \theta \left[
n_B(E,T_\textrm{ph})+1 \right] \left[1 -
\textrm{sinc}(q_{if}R)\right] }{ \pi \rho \hbar c_s \left[(q_{if}
a_{B})^2 + 4 \right] ^4} \,.
\end{equation}
Note that this rate is, in general, a function of the lattice
temperature, $T_\textrm{ph}$, the distance between the donors, $R$
(which fixes $\Delta$ as shown in Fig. \ref{fig:energylevels}),
and the bias between the donors, $\varepsilon$.

\begin{figure}
\centerline{\epsfig{file=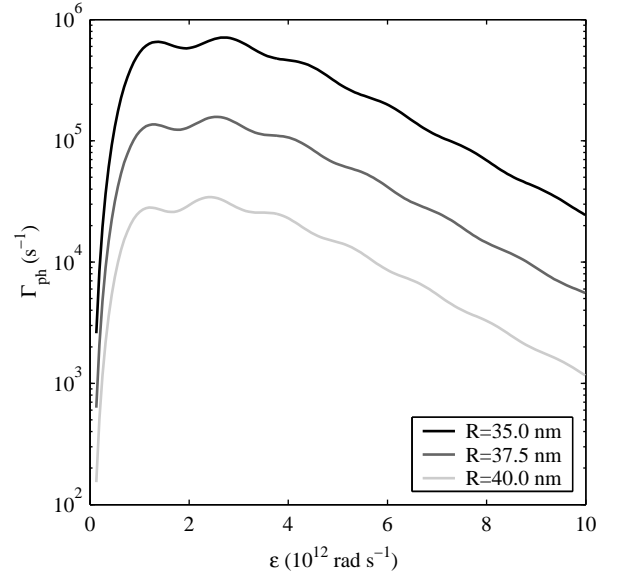}} \caption{Approximate
relaxation rate due to LA phonons, $\Gamma_\textrm{ph}$, as a
function of bias energy, $\varepsilon$, for donor separations $R =
35.0 \textrm{ nm, } R = 37.5 \textrm{ nm, and } R = 40.0 \textrm{
nm}$. We assume $T_\textrm{ph} = 0.1$ K, $D = 3.3$ eV, $\rho =
2.33 \textrm{ kg m}^{-3}$, $c_s = 9.0 \times 10^3 \textrm{m s
}^{-1}$ and $a_0 = 3 \textrm{ nm}$.} \label{fig:phononbias}
\end{figure}
%
% phononbias.eps plotted with H2+\plotrelaxbiasfig.m
%

In Fig. \ref{fig:phononzerobias} we plot $\Gamma_\textrm{ph}$ as a
function of donor separation for zero bias ($\varepsilon = 0$). In
Fig. \ref{fig:phononbias} we plot $\Gamma_\textrm{ph}$ for a
nonzero bias, for three different donor separations ($R = 35.0
\textrm{ nm, } R = 37.5 \textrm{ nm, and } R = 40.0 \textrm{
nm}$). In these calculations we assume $T_\textrm{ph} = 0.1$ K,
$\rho = 2.33 \textrm{ kg m}^{-3}$, $D = 3.3$ eV, $c_s = 9.0 \times
10^3 \textrm{ m s}^{-1}$ (LA phonons) and $a_0 = 3 \textrm{
nm}$.\cite{nsm} Note that the lattice temperature $T_\textrm{ph}$
used here is much less than the effective noise temperature ($T =
10$ K) assumed in Sec. \ref{sec:discussion}; the latter is due to
noise in the electronics used to generate the bias sweep, which is
typically much greater than the sample base temperature.

These results indicate that $\Gamma_\textrm{ph}$ is a strongly
decreasing function of the donor separation (for $R\gtrsim 20$
nm). For donor separations greater than about 35 nm,
$\Gamma_\textrm{ph}^{-1}$ is significantly longer than the other
relevant timescales in the problem, and therefore we are justified
in neglecting phonons as a source of decoherence in this system.

%
%------------------------------------------------------------------------
% REFERENCES
%------------------------------------------------------------------------
%
\bibliography{sepaper}

\end{document}